# The evolutionary history of human populations in Europe

Iosif Lazaridis[1,2]

[1] Department of Genetics, 77 Avenue Louis Pasteur, New Research Building 260, Boston, MA 02115, USA

[2] Broad Institute of Harvard and MIT, Cambridge, Massachusetts 02142, USA

E-mail: lazaridis@genetics.med.harvard.edu

**Abstract**

I review the evolutionary history of human populations in Europe with an emphasis on what has been learned in recent years through the study of ancient DNA. Human populations in Europe ~430-39kya (archaic Europeans) included Neandertals and their ancestors, who were genetically differentiated from other archaic Eurasians (such as the Denisovans of Siberia), as well as modern humans. Modern humans arrived to Europe by ~45kya, and are first genetically attested by ~39kya when they were still mixing with Neandertals. The first Europeans who were recognizably genetically related to modern ones appeared in the genetic record shortly thereafter at ~37kya. At ~15kya a largely homogeneous set of hunter-gatherers became dominant in most of Europe, but with some admixture from Siberian hunter-gatherers in the eastern part of the continent. These hunter-gatherers were joined by migrants from the Near East beginning at ~8kya: Anatolian farmers settled most of mainland Europe, and migrants from the Caucasus reached eastern Europe, forming steppe populations. After ~5kya there was migration from the steppe into mainland Europe and vice versa. Present-day Europeans (ignoring the long-distance migrations of the modern era) are largely the product of this Bronze Age collision of steppe pastoralists with Neolithic farmers.

**Introduction**

The history of human populations in Europe has been studied more extensively than any other continent. Europe is the place where the earliest Neandertal specimens were discovered, pointing to the existence of people in the past who were morphologically distinct from its recent inhabitants. Many other remains were unearthed soon thereafter, and the nascent discipline of physical anthropology was applied to them, inaugurating the scientific, quantitative study of both modern and ancient human populations. During the 20$^{th}$ century anthropology was joined by the study of genetics. The present review focuses on the last few years of this field of study, and in particular on the insights into human history provided by ancient DNA, a powerful new tool in the kit of the prehistorian and evolutionary biologist.

**Archaic Europeans**

The oldest sampled nuclear DNA from Europe dates to ~430kya from Sima de los Huesos in Spain and it was found to be more closely related to Neandertals than to Denisovans[1] ••, unlike mtDNA from the same population, which formed a clade with Denisovan mtDNA[2]. Genome-wide data from several European Neandertal individuals has been published down to ~39kya[3,4]. The Neandertal population that contributed DNA to all non-Africans was more closely related to European Neandertals than to an early (~150kya?) Neandertal from the Altai region in Siberia[3]. Thus, the greater part of the history of European populations was dominated by the Neandertals and their ancestors, raising questions about why this population was replaced in what amounts to a geological blink of an eye.

The genetic divergence between modern humans and Neandertals is lower-bounded by the finding that the Sima de los Huesos hominins already belonged to the Neandertal lineage[1] which must therefore have been already in existence by ~430kya. Modern human and Neandertal Y-chromosomes shared a most recent common ancestor ~450-800kya[5] also pointing to an earlier split of the two lineages. Neandertal mtDNA shared a most recent common ancestor ~270kya, with the earliest known split represented by a specimen from Hohlenstein–Stadel in Germany[6]; the common ancestor of Neandertal and present-day human mtDNA was dated to ~300-500kya[7], appearing to be younger than the corresponding Y-chromosome most recent common ancestor, suggesting that some gene flow may have occurred between the ancestors of modern humans and Neandertals after their separation.

A remarkable recent technical development is the discovery of the fact that mammalian mtDNA is preserved in cave sediments[8] •• and its application to the study of archaic hominins from Europe and Siberia. Sediment DNA allows one to detect the presence of humans in sites where they may be archaeologically invisible and to obtain DNA from the deep past where hominin remains may be scarce or too precious to submit to destructive sampling. A potential limitation is contamination with modern human DNA, which may make this technique more applicable to extinct lineages that could not have plausibly been contributed by modern humans.

**Upper Paleolithic Europeans**

It is only by ~39-36kya that the first sample that clearly shares ancestry with Europeans but not East Asians is evident (Kostenki14 in European Russia[9]), with the earliest known such sample from western Europe at ~35-34kya (GoyetQ116-1 from Belgium[10] ••). Did these and other early Europeans[10,11] • represent a migration into Europe post the Campanian Ignimbrite (CI) volcanic eruption[12] ~39kya, or are they survivors of this event which set off a short period of intense cooling? The ~42-37kya sample from Oase1 in Romania is the earliest known modern human from Europe, and may predate this event. Oase1 has an excess of 6-9% Neandertal ancestry within a genealogical timeframe of 4-6 generations and no specific affinity to Europeans[13], suggesting that at least some of the pre-CI Europeans were replaced after this event. It may be that both the modern human and Neandertal inhabitants of Europe suffered a common demise ~39kya.

Intriguingly, GoyetQ116-1 and Kostenki14—two of the earliest samples on the lineage leading to later Europeans—were not symmetrically related to non-Europeans, with GoyetQ116-1 being genetically closer to a ~40kya sample from China (Tianyuan[14]). In a similar vein, mtDNA haplogroup M, rare in Europe today, but common in eastern non-Africans was present in pre-Last Glacial Maximum Europeans[15]. Surprisingly, this eastern non-African affinity disappeared in samples from the Gravettian-associated "Vestonice-cluster" ~31-26kya which included samples from Italy, Belgium, and the Czech Republic, but partially re-appeared in the ensuing Magdalenian-associated "El Mirón Cluster" ~19-15kya known from sites in Spain, France, Germany, Belgium and Germany[10].

Western European hunter-gatherers (WHG), first described in three sites of western Europe[16-19] are now known to also have lived in southeastern Europe[20,21]•, Switzerland[22], the Baltics[21,23]•, and Italy[10]; the early example from Villabruna in Italy ~15kya [10] has given this population the alternative name "Villabruna cluster". The appearance of WHG ~15kya corresponds to the Bølling-Allerød interstadial warm period, and marked a genetic attraction of European and Near Eastern populations[10,24]••. Was this due to migration between Europe and the Near East during this favorable climatic period or due to the expansion of related populations in Europe and the Near East that had been established there at an earlier period[10]? WHG-like ancestry may represent a partial source of ancestry of populations bordering Europe, for example in Anatolia whose early farmers ~8kya are genetically closer to WHG than other Near Eastern populations are[24], or in the Atlantic where pre-colonial Guanche inhabitants of the Canary Islands had some European hunter-gatherer affinity in addition to their mainly North African origin[25]. WHG did not, however, appear to make any quantifiable genetic contribution to the Upper Paleolithic inhabitants of geographically proximate Morocco ~15kya in North Africa[26].

Eastern European hunter-gatherers (EHG), a population of mixed WHG and Upper Paleolithic Siberian ancestry (related to the Mal'ta and AfontovaGora specimens from Lake Baikal[10,27] ~24-17kya) are attested in European Russia ~8kya[28,29]. This group contributed ancestry to hunter-gatherers in Sweden ~8-5kya[16,30], Norway[31]•, the Balkans and Ukraine[20,21], and the Baltic[21,23,31-33]•. The spread of this ancestry across northern Europe was followed by >3.5kya by the spread of Siberian ancestry[34] that seems to be associated with Finno-Ugrian speakers[16].

**First Farmers**

The dominance of the WHG across much of mainland Europe was relatively short-lived, as they were largely replaced beginning in the 7$^{th}$ millennium BC by farmers from Anatolia via southeastern Europe[29,35]•, who minimally, but variably, mixed with incoming farmers in southeastern Europe[21] and propagated their ancestry as far as Scandinavia[30,36] and Iberia[28,37]. The WHG populations were, however, persistent, with individuals of predominantly WHG ancestry found in early Neolithic contexts in Hungary[19] and as late as the 4$^{th}$ millennium BC in the Blätterhöhle site in Germany[38,39]. These WHG survivors engendered a resurgence of hunter-gatherer ancestry across Middle Neolithic Europe[28] which appears to have involved local populations of hunter-gatherers[38] • rather than a migration from a single WHG-rich area.

Over the Ice Age, Neandertal ancestry seems to have been reduced in Europe[10] by natural selection against Neandertal variants[40,41]; a further reduction was effected during the Neolithic period by migration from the Near East whose populations descended in part from a postulated group of "Basal Eurasians"[16,24], a population that split off from other non-Africans before they split off from each other. Basal Eurasians are consistent with having no Neandertal ancestry at all[24] and may have split from other non-Africans ~101-67kya[42]. It is unknown whether the Basal Eurasians represent descendants of early modern humans in the Near East, or a later entrant into the region. Whatever their origins, both Anatolian farmers and Caucasus hunter-gatherers[22] had ancestry from this population, and through them so do later Europeans via both the Anatolian→Europe and Caucasus→steppe routes discussed below.

**Steppe migrants**

Steppe populations during the Eneolithic to Bronze Age were a mix of at least two elements[28], the EHG who lived in eastern Europe ~8kya and a southern population element related to present-day Armenians[28], and ancient Caucasus hunter-gatherers[22], and farmers from Iran[24]. Steppe migrants made a massive impact in Central and Northern Europe post-5kya[28,43]. Some of them expanded eastward, founding the Afanasievo culture[43] and also eventually reached India[24]. These expansions are probable vectors for the spread of Late Proto-Indo-European[44] languages from eastern Europe into both mainland Europe and parts of Asia, but the lack of steppe ancestry in the few known samples from Bronze Age Anatolia[45] • raises the possibility that the steppe was not the ultimate origin of Proto-Indo-European (PIE), the common ancestral language of Anatolian speakers, Tocharians, and Late Proto-Indo

Europeans. In the next few years this lingering mystery will be solved: either Anatolian speakers will be shown to possess steppe-related ancestry absent in earlier Anatolians (largely proving the steppe PIE hypothesis), or they will not (largely falsifying it, and pointing to a Near Eastern PIE homeland).

Our understanding of the spread of steppe ancestry into mainland Europe is becoming increasingly crisp. Samples from the Bell Beaker complex[46] • • are heterogeneous, with those from Iberia lacking steppe ancestry that was omnipresent in those from Central Europe, casting new light on the "pots vs. people" debate in archaeology, which argues that it is dangerous to propose a tight link between material culture and genetic origins. Nonetheless, it is also dangerous to dismiss it completely. Recent studies have shown that people associated with the Corded Ware culture in the Baltics[23,33] were genetically similar to those from Central Europe and to steppe pastoralists[28,43], and the people associated with the Bell Beaker culture in Britain traced ~90% of their ancestry to the continent, being highly similar to Bell Beaker populations there. Bell Beaker-associated individuals were bearers of steppe ancestry into the British Isles that was also present in Bronze Age Ireland[47], and Iron Age and Anglo-Saxon England[48]. The high genetic similarity between people from the British Isles and those of the continent makes it more difficult to trace migrations into the Isles. This high similarity masks a very detailed fine-scale population structure that has been revealed by study of present-day individuals[49]; a similar type of analysis applied to ancient DNA has the potential to reveal fine-grained population structure in ancient European populations as well.

Steppe ancestry did arrive into Iberia during the Bronze Age[50], but to a much lesser degree. A limited effect of steppe ancestry in Iberia is also shown by the study of mtDNA[51], which shows no detectible change during the Chalcolithic/Early Bronze Age[51], in contrast to central Europe[52]. Sex-biased gene flow has been implicated in the spread of steppe ancestry into Europe[33,53], although the presence and extent of such bias has been debated[54,55]. One aspect of the demographies of males and females was clearly different, as paternally-inherited Y-chromosome lineages experienced a bottleneck <10kya which is not evident in maternally-inherited mtDNA[56], suggesting that many men living today trace their patrilineal ancestry to a relatively small number of men of the Neolithic and Bronze Ages.

Southeastern Europe received steppe-related ancestry before any other population in Europe outside the steppe itself, with sporadic appearance of individuals with steppe ancestry in Bulgaria as early as ~6.7-6.5kya[21] and a general low-level presence of ~30% during the Bronze Age, ~5.4-3.1kya. This ancestry was also present in the Aegean during the Mycenaean period ~3.5kya at ~15%, but was absent from the otherwise genetically similar Minoan culture of Crete who represents the most recent sampled European population without any such ancestry[45]. Both Minoans and Mycenaeans[45], and to a much lesser extent Neolithic samples from the Peloponnese and Bulgaria also had ancestry related to Caucasus hunter-gatherers, suggesting that this ancestry did not come to Europe only via migrations from the steppe, but also independently, perhaps reflecting ancestry from different Anatolian source populations[29,35,57,58] •.

**The future of Europe's past**

Europe's role in world history is often rightfully viewed today through the lens of the impact of its recent colonial past on indigenous populations around the world. Yet ancient DNA is revealing that similar processes of migration, conflict, extinction, and occasional peaceful co-existence played out within the borders of Europe itself.

Ancient DNA research has helped us reconstruct many previously unsuspected events in European evolutionary history (Fig. 1), but it also opened up many new questions: Where did the first European-like humans (represented by Kostenki14 and GoyetQ116-1) arrive from? What contributed to the genetic differentiation between Europe and the Near East; how did Near Eastern populations acquire Basal Eurasian ancestry and why did this ancestry not spread into geographically neighboring Europe? How did the WHG become so successful in Europe ~15kya virtually replacing all previous Ice Age Europeans? What drove ancient Siberian populations westward to produce the European/Siberian intermediate EHG? What was the mechanism by which steppe populations of mixed EHG and Near Eastern ancestry were formed, and what was the mechanism by which steppe populations managed to make a major demographic impact in Neolithic Europe? The later history of European population will probably be no less interesting: how did the elements of the populations of Europe, mostly present in the continent since the Bronze Age, combine during the Migration Period[59,60] and the centuries that followed to form the remarkable genetic, linguistic, and cultural diversity of present-day Europeans?

## Acknowledgments

I thank E. Harney, M. Lipson, I. Mathieson, V. Narasimhan, and D. Reich for comments on the paper.

## Conflict of Interest

The author declares no conflict of interest.

**Figure 1: A sketch of European evolutionary history based on ancient DNA.** Bronze Age Europeans (~4.5-3kya) were a mixture of mainly two proximate sources of ancestry: (i) the Neolithic farmers of ~8-5kya who were themselves variable mixtures of farmers from Anatolia and hunter-gatherers of mainland Europe (WHG), and (ii) Bronze Age steppe migrants of ~5kya who were themselves a mixture of hunter-gatherers of eastern Europe (EHG) and southern populations from the Near East. Thus, we only have to go ~8 thousand years backwards in time to find at least four sources of ancestry for Europeans. But, each of these sources was also admixed: European hunter-gatherers received genetic input from Siberia and ultimately also from archaic Eurasians, and Near Eastern populations interacted in unknown ways with Europe and Siberia and also had ancestry from 'Basal Eurasians', a sister group of the main lineage of all other non-African populations. Dates correspond to sampled populations; in the case of a cluster of populations (such as the WHG), they correspond to the earliest attestation of the group.

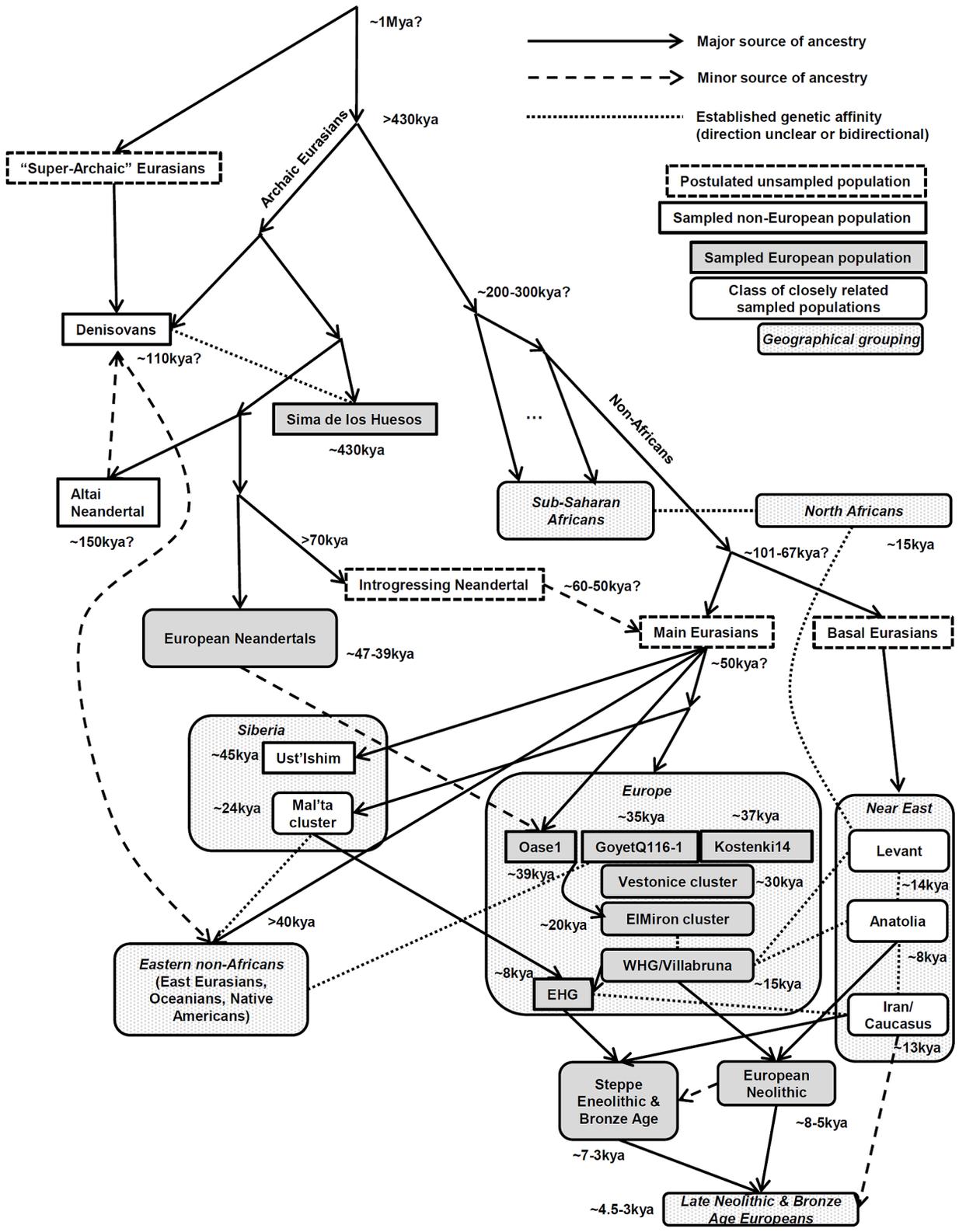